\documentclass[twocolumn,showpacs,preprintnumbers,amsmath,amssymb,pra,epsf]{revtex4}
\usepackage{graphicx}
\begin{document}
\title{Electronic noise-free measurements of squeezed light}

\author{Leonid A. Krivitsky$^{1,2}$, Ulrik L. Andersen$^2$, Ruifang Dong$^1$, Alexander Huck$^2$, Christoffer Wittmann$^1$, and Gerd Leuchs$^1$}

\affiliation{$^1$ Institute of Optics, Information and Photonics,
Max-Planck Research Group, University Erlangen-Nuremberg, G. Scharowsky street 1, 91058 Erlangen, Germany\\
$^2$ Technical University of Denmark, Department of Physics,
Fisikvej 309, 2800 Lyngby, Denmark}

\begin{abstract}
\begin{center}\parbox{14.5cm}
{We study the implementation of a correlation measurement technique
for the characterization of squeezed light. We show that the sign of
the covariance coefficient revealed from the time resolved
correlation data allow us to distinguish between squeezed, coherent
and thermal states. In contrast to the traditional method of
characterizing squeezed light, involving measurement of the
variation of the difference photocurrent, the correlation
measurement method allows to eliminate the contribution of the
electronic noise, which becomes a crucial issue in experiments with
dim sources of squeezed light.}
\end{center}
\end{abstract}
\maketitle \narrowtext
\vspace{-10mm}

The pioneering experiments of Hanbury-Brown and Twiss ~\cite{twiss}
on the implementation of the intensity interferometer, based on
correlation measurements, offered great potentials in modern optics.
Nowadays, this technique plays an essential role in the
\emph{discrete variables quantum optics}, where correlated photons
are generated in various nonlinear optical processes. The
observation of the simultaneous detection events (coincidences) from
spatially separated single-photon detectors reveals the nonclassical
correlation properties of the system. Such a correlation measurement
strategy is indispensable in many quantum optical experiment such as
tests of the foundations of quantum mechanics~\cite{chuang}, quantum
cryptography~\cite{gisin}, and quantum metrology ~\cite{alan}.

In \emph{continuous variables quantum optics}, associated with the
manipulation and detection of a continuous degree of freedom, a
different measurement technique is traditionally implemented. The
most common method here is balanced homodyne detection (HD), where
the squeezed light is mixed on a symmetric beamsplitter with an
auxiliary beam, denoted the local oscillator (LO) and the difference
photocurrent of two analogue photodetectors is recorded
~\cite{review sqz} (Fig.1). This method represents the basis of
optical homodyne quantum tomography, thus facilitating a complete
reconstruction of the Wigner function. In fact, the potential of the
HD has been proven in many experiments with continuous variable
systems ~\cite{raymer} and also in experiments with their discrete
variables counterparts ~\cite{lvovsky, bellini}. However, the
realistic HD setup suffers from optical losses and electronic noise
(EN) of the detectors. These factors mask the nonclassical
properties of the detected light and therefore limits the
performance of the HD. However, detailed knowledge of the statistics
of the optical loss and the EN allows to account for these effects,
using special
mathematical algorithms~\cite{raymer,lvovskyelnoise}. 
As an alternative to this inference method one can directly measure
the optical variance free from the EN using a correlation
measurement method, commonly used in discrete variable quantum
optics. Such a detector noise suppression technique has been applied
to other areas~\cite{ziel} and its application to continuous
variable quantum optics was proposed in~\cite{leuchs}. Therefore,
the present paper reports its experimental demonstration.

Characterization of squeezed light by correlation measurements was first theoretically
proposed in~\cite{hom}. Later, the correlation function of the output of the optical
parametric amplifier was experimentally measured ~\cite{kono}, where direct access to the photon number
correlations was obtained using single photon detectors. An alternative approach was developed in ~\cite{pklam},
 where correlations between different quadratures in HD setup were used to study the photon statistics of the optical parametric oscillator.

Let us consider a standard HD setup as shown in Fig.1. We express
the field operator as $a=\alpha+\delta{a}$, where $\alpha$
represents the "classical" bright component and $\delta{a}$ is an
operator with zero mean value, describing the quantum fluctuations
of the field amplitude. It is useful to introduce also the amplitude
and phase quadratures, which are given by $X=1/2(a^\dagger+a)$ and
$Y=i/2(a^\dagger-a)$, respectively. We consider the input state to
be a \emph{squeezed vacuum} state. Following the standard
formalism~\cite{bachorbook}, and accounting for the EN, we derive
the photocurrents of two detectors in each port of the beamsplitter:
\begin{equation}
i_{1,2}=1/2\alpha^2_{LO}+\alpha_{LO}(\delta{X_{LO}}\pm\delta{X_{\phi}})+\delta{i_{el1,2}}\\
\end{equation}
where $\alpha_{LO}$ is a mean field amplitude of the LO, $\delta
X_\phi\equiv\cos{\phi}\delta X+\sin{\phi}\delta Y$ is an operator
representing the quantum fluctuations of the tilted quadrature of
the input beam, $\delta{X_{LO}}$ is the amplitude quadrature of the
LO and $\delta{i_{el1,2}}$ are stochastic numbers associated with
the EN. In order to reveal the squeezing properties of the input
beam, the variance of the difference photocurrent is traditionally
recorded (onwards referred to as \emph{the substraction method}).
One obtains the following expression:
\begin{equation}
\langle(i_{1}-i_{2})^2\rangle=4\alpha^2_{LO}\langle\delta{X^2_{\phi}}\rangle+\langle\delta{i^2_{el1}}\rangle+\langle\delta{i^2_{el2}}\rangle.
\end{equation}
From (2) it is clear, that the contribution from the EN always
affects the substraction method. Thus, in case if the EN is
prevailing over the optical signal, the measurement data would
mainly represent EN. We also note that the shot noise level (SNL)
can be calibrated by blocking the input signal state, so that
$\delta X_{\phi}\equiv \delta X_{vac}$ assuming
$4\alpha^2_{LO}>>\langle\delta{i^2_{el1}}\rangle+\langle\delta{i^2_{el2}}\rangle$.

\begin{figure}
\includegraphics[width=8cm]{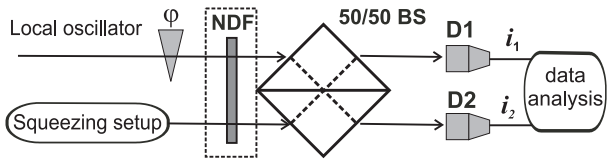}
\caption{The scheme of the HD setup. The beam from the squeezing
setup is interfering with the LO on a 50/50 beamsplitter
\textbf{BS}, $\phi$ is the phase of the LO, \textbf{D1} and
\textbf{D2} are analogue photodetectors. Linear attenuation is
introduced in both beams by the neutral density filter \textbf{NDF}.
The produced photocurrents $i_1$ and $i_2$ are further analyzed by
the computer.}
\end{figure}

As an alternative to the subtraction method we now investigate the
correlation method. We obtain the following expression for the
covariance coefficient of two photocurrents
\begin{equation}
\hbox{cov}(i_{1},i_{2})\equiv\langle{i_{1}i_{2}}\rangle-\langle{i_{1}}\rangle\langle{i_{2}}\rangle=\alpha^2_{LO}[\langle\delta{X^2_{LO}}\rangle-
\langle\delta{X^2_{\phi}}\rangle],
\end{equation}
where it is assumed that the EN of two detectors are not correlated,
i.e.
$\langle\delta{i_{el1}}\delta{i_{el2}}\rangle=\langle\delta{i_{el1}}\rangle\langle\delta{i_{el2}}\rangle$.
As it is seen from (3) the covariance coefficient is completely
independent on the EN due to the time averaging of the data and
statistical independence of the noises of two photodetectors. For
the sake of clarity, we now assume that the beam of the LO is shot
noise limited, i.e. $\langle\delta{X^2_{LO}}\rangle=
\langle\delta{X^2_{vac}}\rangle$. The analysis of the formula (3)
suggests, that the sign of the covariance is determined by the
statistics of the incoming light. Indeed, if the input state is
squeezed, i.e. $\langle\delta{X^2_{\phi}}\rangle<
\langle\delta{X^2_{vac}}\rangle$, then the covariance is positive.
In contrast, if the input state is coherent, i.e.
$\langle\delta{X^2_{\phi}}\rangle= \langle\delta{X^2_{vac}}\rangle$
then the covariance equals zero, and finally, if the input state
exhibits classical excess noise, i.e.
$\langle\delta{X^2_{\phi}}\rangle> \langle\delta{X^2_{vac}}\rangle$,
the covariance is negative. Therefore, the sign of the covariance
coefficient allows one to distinguish between different kinds of
input states independently on the amount of EN, provided that one
has information about the noise of the LO.

In our experiment we used a squeezing source based on the Kerr
nonlinearity in an optical fiber, which is pumped by a femtosecond
pulsed laser at the wavelength of $1.5\mu m$. Combining two
 orthogonally polarized quadrature squeezed beams with a fixed
relative phase, we generated a special kind of two mode squeezed
states (known as polarization squeezing)~\cite{polsqueezing}, where
the squeezed beam and the beam of the LO have orthogonal
polarizations and propagate in the same spatial mode. Therefore, the
interference between the LO and the squeezed pulses is achieved by
interfering them at the polarizing beamsplitter with a relative
phase controlled by a half-wave plate. Schematically, our
experimental setup is completely equivalent to the one shown in
Fig.1, where the squeezed pulses and the LO propagate in different
spatial modes and interfere on a symmetric beamsplitter. The two
output beams of the beamsplitter are measured with PIN diodes. The
phase of the local oscillator as well as the total attenuation of
the two input beams were adjustable, as shown in the figure. In each
experimental run we recorded the raw data from the two detectors by
digitizing the AC components of the photocurrents. For this purpose
we used a high-speed digitizer with a sampling rate of
$20*10^6\hbox{samples/s}$ whilst the bandwidth of the detected
signal was restricted by the low-pass RF-filter with a full width at
half maximum of $3\hbox{MHz}$. We applied a simple Mathlab script to
one and the same recorded data in order to calculate the covariance,
and for comparison, the variance of the difference photocurrent.

We performed two series of measurements: the first one aimed at
witnessing the squeezing and anti-squeezing by correlation
measurements and the second one aimed at testing of the EN-free
detection of the degree of squeezing.

{\it Witnessing squeezing:} The quadrature of the squeezed beam
being observed by the detectors is controlled by the relative phase,
$\phi$, of the LO. Thus the phase of the LO determines whether
correlation or anti-correlation between the detectors is measured.
Measurements of the correlation coefficient through a scan of the LO
is shown in Fig.2. In this measurement run the average power of the
LO was set to $6\hbox{mW}$, providing a $17\hbox{dB}$ clearance
between the EN level and the SNL. The zero covariance was observed
with the coherent beam of the same intensity. The positive
covariance corresponds to the measurement of the squeezed
quadrature, whilst the negative covariance corresponds to the
measurement of the anti-squeezed quadrature. Note, that the modulus
of the positive covariance values are much smaller than the negative
ones, which suggests that the measured quantum state was not pure.
The impurity arises from various linear and nonlinear effects in the
optical fiber~\cite{polsqueezing2}.

\begin{figure}
\includegraphics[width=8.5cm]{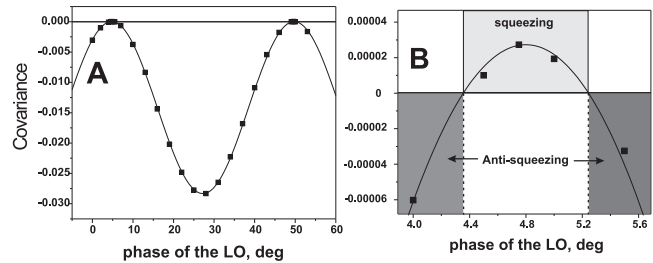}
\caption{Experimental dependence of the covariance versus the phase
of the LO. (A) is the dependence on the full range of the phase
variation and (B) is the magnified part of the dependence with the
positive covariance. Light-gray shaded region corresponds to the
measurement of the squeezed quadrature, whilst the dark-grey shaded
region corresponds to the measurement of the quadrature exhibiting
excess noise. The line represents the fit of experimental data.}
\end{figure}

{\it Electronic noise free detection:} In the second series of
measurements we investigate the influence of the measured degree of
squeezing with attenuation of the signal and LO by a neutral density
filter (NDF). In case of a strong attenuation, the produced
photocurrents carry a considerable amount of the EN relative to the
optical noise. As mentioned above, this EN affects the subtraction
method but it should not appear when the correlation coefficient is
measured. We adjust the phase of the LO such as to minimize the
detected noise variance (corresponding to a measurement of the
squeezed quadrature). By using the standard subtraction method we
obtain results of the degree of squeezing summarized in Fig.3A. The
variances of the measured squeezed quadrature are normalized to the
EN free SNL. To establish an EN free calibration of the SNL we
perform noise measurements of a shot noise limited LO at very high
powers where the optical shot noise dominates over the EN. Knowing
the linear behavior of the shot noise, we subsequently extrapolated
the data to yield a reliable (thus EN free) calibration of the SNL
for all powers of the LO. From Fig.3A, we see that at maximum
optical power, when the influence of the EN is negligible, we
observed $-1.65\hbox{dB}$ of squeezing. However, when the beams are
strongly attenuated the EN starts to play a role and at a certain
attenuation (about 80\%), excess noise is observed thus indicating
the dominance of the EN. For comparison, we also plot the expected
squeezing of the state as it would be measured with an ideal EN-free
detector as a function of the attenuation (solid traces in Fig.3).

\begin{figure}
\includegraphics[width=8.5cm]{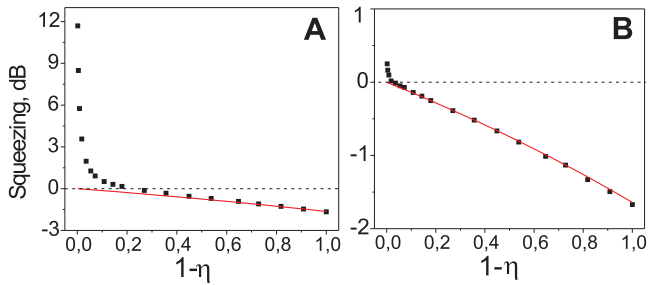}
\caption{Experimental dependence of the squeezing on the
transmission $1-\eta$:(A) for the substraction method and (B) for
the covariance method. Red solid line represents the theoretical
dependence without accounting for the EN. Horizontal dashed line at
zero defines a border between the squeezing and the classical access
noise.}
\end{figure}

We now use the correlation method to find the variance of the
squeezed quadrature. After measuring the correlation coefficient,
the variance is found from (3) with the shot noise calibration being
identical to the one used above for the subtraction method. The
results are presented in Fig.3B, where we also include the
theoretically expected variances for noise-free detection (solid
line). The theoretical curve is seen to fit the experimental data up
to very low powers. Therefore, in contrast to the subtraction
method, we see that the correlation method is less affected by the
EN, even when the optical power is very low.

In the experiment we observe a small but non-zero background
correlation coefficient stemming from the EN of the detectors, which
turned out to be slightly correlated. Such a background correlation
limits the ability to resolve squeezing at very low powers as seen
from the discrepancy at high attenuations in Fig.3B. The correlation
of the EN can be reduced by using completely independent acquisition
systems for the two detectors unless the reason for this correlation
is common pick-up of external electromagnetic signals.

In conclusion, we have experimentally realized a correlation
measurement strategy that yields results that are free of EN of the
detector. Using this method, we have witnessed the presence of
squeezed noise and excess noise, and more importantly we have
performed an EN free detection of squeezed light. The method is
therefore a very interesting alternative to standard homodyne
detector where EN may affect the measurements.

The authors appreciate the discussions with M. Chekhova, R. Filip,
and Ch. Marquardt. This work was supported by the EU project COMPAS and
the Danish Research Council (FTP). One of us
(L.A.K.) acknowledges the support of the Alexander von Humboldt foundation.

\end{document}